%% file: paper.tex
\documentclass[11pt,a4paper]{article}
\usepackage[top=1.1in, bottom=1.1in, left=1.1in, right=1.1in]{geometry}
\usepackage{graphicx}
\usepackage{rotating}
\usepackage{algpseudocode}
\usepackage{stmaryrd}
\usepackage{amsmath}
\usepackage{amssymb}
\usepackage{makeidx}  
\usepackage{url}
\usepackage{ntheorem}
\newcommand{\ignoreme}[1]{}
\newcommand{\tick}{\checkmark}
\newtheorem{definition}{Definition}
\newtheorem{theorem}{Theorem}
\newtheorem{lemma}{Lemma}
\newtheorem{proof}{Proof}
\begin{document}
\title{Heuristics for Selecting Predicates for Partial Predicate Abstraction}
%
%

\author{Tuba Yavuz \\ 
University of Florida\\
tuba@ece.ufl.edu\\
\and 
Chelsea Metcalf\\
University of Florida\\
chelseametcalf@ufl.edu\\
}

\maketitle              

\begin{abstract}
In this paper we consider the problem of configuring partial predicate abstraction 
that combines two techniques that have been effective in 
analyzing infinite-state systems: predicate abstraction and fixpoint approximations. 
A fundamental problem in partial predicate abstraction is deciding the variables to be abstracted and the predicates to be used.
In this paper, we consider systems modeled using linear integer arithmetic and investigate an alternative approach to counter-example guided abstraction refinement. 
We devise two heuristics that search for predicates that are likely to be precise. The first heuristic performs the search on the problem instance to be verified.
The other heuristic leverages verification results on the smaller instances of the problem. We report experimental results for CTL model checking and discuss  advantages and disadvantages of each approach.
\end{abstract}
\section{Introduction}
\input{intro.tex}
\section{Preliminaries}
\label{sec:prel}

\input{prelim.tex}

\section{Choosing the Predicates}
\label{sec:approach}
\input{approach.tex}

\section{Experiments}
\label{sec:exp}
\input{experiments.tex}

\section{Related Work}
\label{sec:relwork}

\input{relWork.tex}

\section{Conclusion}
\label{sec:conc}
\input{conclusions.tex}
\bibliographystyle{abbrv}
\bibliography{cav2016}
\end{document}

%% file: intro.tex
Partial predicate abstraction \cite{Yav16}  is a hybrid technique that combines advantages of predicate abstraction and fixpoint approximations for model checking infinite-state systems. It provides scalability by mapping part of the state space to a set of boolean variables representing a set of predicates on the mapped domain. It provides precision by representing the other part of the state space in its concrete domain and using approximation techniques to achieve convergence for computing the fixpoint. However, as in traditional predicate abstraction, a fundamental problem remains to be addressed: {\em choosing the right set of predicates}.

Counter-example guided abstraction refinement (CEGAR) addresses this fundamental problem in an incremental way. It basically expands the predicate set by discovering new predicates from the point where the spurious counter-example and the concrete behavior diverge. Craig interpolation \cite{Crai57} has been successfully used to automatically compute new predicates that would avoid the spurious counter-example. However, when the analysis continues with the new set of predicates new spurious behavior may be discovered and  new predicates are generated and so on. Since model checking infinite-state systems is undecidable, the CEGAR loop may not terminate. Even if it terminates, it may end up generating a large set of predicates. 

In this paper, we investigate an alternative approach to CEGAR for partial predicate abstraction. Since partial predicate abstraction is capable of 
representing some part of the state space in a concrete way, we need to identify how to partition the state space in terms of abstracting and approximating. We also would like to avoid two major problems with CEGAR: 1) choosing the initial set of predicates on variables that better be kept in their concrete domains and 2) ending up with a large set of predicates.  
We present two approaches to selecting predicates that would yield a precise and a feasible analysis. 
One approach measures the relative imprecision of the candidate predicates on the problem instance to be verified.
Another approach uses a smaller instance of the problem and uses the verification results to infer a precise set of predicates to be used for verifying a larger instance.
 
 Our heuristics are based on the properties of {\em incremental abstraction} that enables sound elimination of some of the candidates. We show that the elimination can be 
 applied in both of the presented approaches. We have implemented both approaches using the Action Language Verifier (ALV) \cite{YB09}, which was also used to  perform the verification experiments. 
 For both approaches we present  algorithms that select an optimal set of predicates based on the elimination heuristics and within the specified bound for the number of predicates.  We use models on Airport Ground Traffic Control and a character-special device driver and perform measurements on various safety and liveness problem instances. 
 Experimental results show effectiveness of the heuristics. We discuss the advantages and disadvantages of each approach.
 
The rest of the paper is organized as follows. Section \ref{sec:prel} presents basic definitions and properties of partial predicate abstraction. Section \ref{sec:approach} presents the two approaches to selecting the appropriate set of predicates.  Section \ref{sec:exp} presents experimental results. Section \ref{sec:relwork} discusses related work and Section \ref{sec:conc} concludes with directions for future work.

%% file: prelim.tex
In this section we present preliminaries on the partial-predicate abstraction technique, which combines predicate abstraction and fixpoint approximations.
We consider transition systems that are described in terms of boolean and unbounded integer variables. 

\begin{definition}
An infinite-state transition system is described by a Kripke structure $T=(S, I, R, V)$, where $S$, $I$, $R$, and $V$ denote the state space, set of initial states, the transition relation, and the set of state variables, respectively. $V=V_{bool} \cup V_{int}$ such that $S \subseteq \mathcal{B}^{|V_{bool}|} \times \mathcal{Z}^{|V_{int}|}$, $I \subseteq S$, and $R \subseteq S \times S$.
\end{definition}

\begin{definition}
Given a Kripke structure, $T=(S, I, R, V)$ and a set of states $A \subseteq S$, 
the pre-image operator, $pre[R](A)$, computes the set of states that can reach the states in $A$  in one step: 
$pre[R](A)=\{ b \ | \ a \in A \ \wedge \ (b,a) \in R \}.$
\end{definition}

\begin{figure}[th!]
\centering
\begin{tabular}{cl} \hline 
$V$. & $s$, $t$, $a_1$, $a_2$, $z$: integer \\
                      & $pc1$, $pc2$: {think, try, cr} \\ 
$I$ & $s=t \wedge pc_1=think \wedge pc_2=think$ \\ 
Transitions: &  \\ 
  $r^{try}_i $ & $\equiv pc_i=think \wedge a_i'= t \wedge t'=t+1 \wedge pc_i'=try$ \\
  $r^{cr}_i $ & $\equiv pc_i=try \wedge s \geq a_i \wedge z'=z+1 \wedge pc_i'=cr $\\
  $r^{think}_i $ & $\equiv pc_i=cr \wedge s'=s+1 \wedge z'=z-1 \wedge pc_i'=think$ \\
$R$: & $\bigvee_{i=1,2} r^{try}_i \vee r^{cr}_i \vee r^{think}_i$ \\ \hline 
\end{tabular}
\caption{The ticket mutual exclusion algorithm for two processes. Variable $z$ is an addition to demonstrate the utilization of the proposed approach.}
\label{fig:ticket}
\end{figure}

\paragraph{Abstract Model Checking and Predicate Abstraction}

\begin{definition}
Let $\varphi$ denote a set of predicates over integer variables. Let $\varphi_i$ denote a member of $\varphi$ and $b_i$ denote the fresh
boolean variable that corresponds to $\varphi_i$. $\bar{\varphi}$ represents an ordered sequence (from index 1 to $|\varphi|$) of predicates in $\varphi$. The set of variables that appear in $\varphi$ is denoted by $V(\varphi)$. Let $\varphi'$ denote the set of next state predicates obtained from $\varphi$ by replacing variables in each predicate $\varphi_i$ with their primed versions. Let $b$ denote the set of $b_i$ that corresponds to each $\varphi_i$. Let $V_{\sharp} = V_{\natural} \cup b \setminus V(\varphi)$\footnote{We will use $\natural$ to refer to the abstracted system and $\sharp$ to refer to the abstract system.}.  
\end{definition}

\paragraph{Abstracting states} A concrete state $s^{\natural}$ is predicate abstracted using a mapping function $\alpha$ via a set of predicates $\varphi$ by introducing a predicate boolean variable $b_i$ that represents  predicate $\varphi_i$ and existentially quantifying the concrete variables $V(\varphi)$ that appear in the predicates:

\begin{equation}
\label{eq:alpha}
\alpha(s^{\natural}) = \exists V(\varphi). (s^{\natural} \ \wedge \ \bigwedge^{|\varphi|}_{i=1} \varphi_i \iff b_i ).
\end{equation}

\paragraph{Concretization of abstract states} An abstract state $s^{\sharp}$ is mapped back to all the concrete states it represents by replacing each predicate boolean variable $b_i$ with the corresponding predicate $\varphi_i$: 
\begin{equation}
\label{eq:gamma}
\gamma(s^{\sharp}) = s^{\sharp} [\bar{\varphi}/\bar{b}]
\end{equation}

\ignoreme{
Abstraction function $\alpha$ provides a safe approximation for states:

\begin{lemma}
\label{lemma:galoisState}
$(\alpha, \gamma)$, as defined in Equations \ref{eq:alpha} and \ref{eq:gamma}, defines a Galois connection, i.e., $\alpha$ and $\gamma$ are monotonic functions and $s^{\natural} \subseteq \gamma(\alpha(s^{\natural}))$ and $\alpha(\gamma(s^{\sharp}))=s^{\sharp}$ (see the Appendix for the proof).
\end{lemma}
}

A concrete transition system can be conservatively approximated by an abstract transition system through a simulation relation  or a surjective mapping function involving the respective state spaces in terms of existential abstraction. 

\begin{definition} (Existential Abstraction)
Given transition systems $T_1=(S_1, I_1, R_1, V_1)$ and $T_2=(S_2,I_2,R_2,V_2)$, $T_2$ approximates $T_1$ (denoted $T_1 \sqsubseteq_{h} T_2$) iff
\begin{itemize}
\item $\exists s_1. (h(s_1) = s_2 \ \wedge \ s_1 \in I_1)$ implies $s_2 \in I_2$,
\item $\exists s_1, s_1'. (h(s_1)=s_2 \ \wedge \ h(s_1')=s_2' \ \wedge \ (s_1,s_1') \in R_1)$ implies $(s_2, s_2') \in R_2$,
\end{itemize} 
where $h$ is a surjective function from $S_1$ to $S_2$.
\end{definition}

It is a known  \cite{Loi95} fact that one can use a Galois connection $(\alpha, \gamma)$\footnote{$\alpha(c) \sqsubseteq a \text{ iff } c \sqsubseteq \gamma(a)$, where 
concrete value $c$ maps to abstract value $a$ through $\alpha$. Here we interpret $\sqsubseteq$ as the logical implication operation.} to construct an approximate transition system. Basically, $\alpha$ is used as the mapping function and $\gamma$ is used to map properties of the approximate or abstracted system to the concrete system.

\begin{theorem}
\label{theorem:amc}
Assume $T_1 \sqsubseteq_{\alpha} T_2$, $\phi$ denotes an ACTL\footnote{ACTL is fragment of CTL that involves temporal operators that preceded by universal quantification, $A$, over states} formula that describes a property of $T_2$, $C(\phi)$ denotes the transformation of the correctness property
by descending on the subformulas recursively and transforming every  atomic formula $a$ with $\gamma(a)$ (see \cite{CGL94} for details), and $(\alpha,\gamma)$ forms a Galois connection and defines predicate abstraction and concretization as given in Equations \ref{eq:alpha} and \ref{eq:gamma}, respectively. Then, $T_2 \models \phi$ implies $T_1 \models C(\phi)$ (see \cite{Yav16} for the proof).
\end{theorem}

For example, let $\phi$ be $AG(b_1 \vee b_2)$, where $b_1$ and $b_2$ represent $z=1$ and $z<1$, respectively,  when the model in Figure \ref{fig:ticket} is predicate abstracted wrt to the set of predicates $\varphi=\{z=1, z<1\}$ and the Galois connection  $(\alpha, \gamma)$ defined as in Equations \ref{eq:alpha} and \ref{eq:gamma}. Then, $C(\phi)=AG(z\leq1)$.

\subsubsection{Computing A Partially Predicate Abstracted Transition System}
\label{sec:partial}
We compute an abstraction of a given transition system via a set of predicates such that only the variables that appear in the predicates disappear, i.e., existentially quantified, and all the other variables are preserved in their concrete domains and in the exact semantics from the original system. As an example, using the set of predicates $\{z=1, z<1\}$, we can partially abstract the model in Figure \ref{fig:ticket} in a way that $z$ is removed from the model, two new boolean variables $b_1$ (for $z=1$) and $b_2$ (for $z<1$) are introduced, and $s$, $t$, $a_1$, $a_2$, $pc_1$, and $pc_2$ remain the same as in the original model.

\paragraph{Abstracting transitions} A concrete transition $r^{\natural}$ is predicate abstracted using a mapping function $\alpha^{\tau}$ via a 
set of current state  predicates $\varphi$ and a set of next state  predicates $\varphi'$  by introducing a predicate boolean variable $b_i$ that represents predicate $\varphi_i$ in the current state and a predicate boolean variable $b_i'$ that represents predicate $\varphi_i$ in the next state and existentially quantifying the current and next state concrete variables $V(\varphi) \cup V(\varphi')$ that appear in the current state and next state predicates.

\begin{equation}
\label{eq:alphaT}
\alpha^\tau(r^{\natural}) = \exists V(\varphi). \exists V(\varphi'). (r^{\natural} \ \wedge \ CS \ \wedge \ \bigwedge^{|\varphi|}_{i=1} \varphi_i \iff b_i \wedge \ \bigwedge^{|\varphi|}_{i=1} \varphi_i' \iff b_i'  ),
\end{equation}

where $CS$ represents a consistency constraint that if all the abstracted variables that appear in a predicate remains the same in the next state then the corresponding boolean variable is kept the same in the next state:

\[CS = \bigwedge_{\varphi_i \in \varphi} ((\bigwedge_{v \in V(\varphi_i)} v'=v ) \implies b_i' \iff b_i).\] 

\ignoreme{
\paragraph{Concretization of abstract transitions} An abstract transition $r^{\sharp}$ is mapped back to all the concrete transitions it represents by replacing each current state boolean variable $b_i$ with the corresponding current state predicate $\varphi_i$ and each next state boolean variable $b_i'$ with the corresponding next state predicate $\varphi_i'$:

\[\gamma^{\tau}(r^{\sharp}) = r^{\sharp} [\bar{\varphi},\bar{\varphi}'/\bar{b},\bar{b}']\]
}
As an example, for the predicate set $\{a>b, c=0\}$, $CS \equiv ((a'=a \wedge b'=b \implies b_1'=b_1) \wedge (c'=c \implies b_2'=b_2))$, where $b_1$ represents predicate $a>b$ and $b_2$ represents predicate $c=0$.

For the model in Figure \ref{fig:ticket} and predicate set $\phi=\{z=1, z<1\}$, partial predicate abstraction of $r^{cr}_i$, $\alpha^{\tau}(r^{cr}_i)$, is  computed as 
$pc_i= try \ \wedge \ s \geq a_i \ \wedge \ ((b_1 \wedge \neg b_2  \wedge \neg b'_1  \wedge  \neg b'_2)  \vee \ (\neg b_1 \wedge  b_2 \wedge  (b'_1  \vee  b'_2))$ $
 \vee \ (\neg b_1 \wedge \neg b_2 \wedge \neg b'_1 \wedge \neg b'_2))   \ \wedge \ pc_i'=cr.$

It is important to note that the concrete semantics pertaining to the integer variables $s$ and $a_i$ and the enumerated variable $pc_i$ are preserved in the partially abstract system.
 
 The main merit of the combined approach is to combat the state explosion problem in the verification of problem instances  for which predicate abstraction does not provide the necessary precision (even in the case of being embedded in a CEGAR loop) to achieve a conclusive result.  
 As we have shown in \cite{Yav16}, in such cases  approximate fixpoint computations \cite{YB09} may turn out to be more precise. The hybrid approach may provide both the necessary precision to achieve a conclusive result and an improved performance by predicate abstracting the variables that do not require fixpoint approximations.  

\subsection{Counter-example Guided Abstraction Refinement for Partial Abstraction}
\label{sec:cegaar}

A common approach for dealing with imprecision in predicate abstraction is Counter-Example Guided Abstraction Refinement (CEGAR). The idea is to analyze the spurious counter-example path, identify the cause of divergence between the concrete behavior and the abstract path, and refine the system by extending the predicate set with refinement predicates. 
One of the challenges in CEGAR is controlling the size of the predicate set as new predicates get added. Since predicate abstracting a transition system is exponential in the number of predicates, if not controlled, CEGAR can easily blow up before providing any conclusive result.

We have implemented CEGAR for partial predicate abstraction for ACTL model checking (see \cite{Yav17a} for details), which we will refer as CEGAAR in the rest of the paper\footnote{The phrase stems from the fact that both abstraction and approximation, the two As, are guided by counter-examples.}.
To deal with the predicate set size, we have used a breadth-first search (BFS) strategy to explore the predicate choices until  a predicate set  producing a conclusive result can be found. The algorithm keeps a queue of sets of predicates  and in each iteration it removes a predicate set from the queue and computes the partial predicate abstraction with that predicate set. When a set of refinement predicates $rp$ is discovered for a given spurious counter-example path, rather than extending the current predicate set $\varphi$ with $rp$ in one shot, it considers as many extensions of $\varphi$ as $|rp|$ by extending $\varphi$ with a single predicate from $rp$ at a time and adds all these predicate sets to the queue to be explored using BFS. In the context of partial predicate abstraction, this strategy has been more effective in generating conclusive results compared to adding all refinement predicates at once.

%% file: approach.tex
In this section, we present two approaches to choosing predicates for partial predicate abstraction. Both approaches build on  the concept of incremental abstraction and the guarantees provided by such abstractions that guide elimination of the candidate predicates. So, first we introduce the concepts related to incremental abstraction in Section \ref{sec:threory} and present the individual approaches in sections \ref{sec:onsame} and \ref{sec:small}.

\ignoreme{We need an oracle to tell us whether a predicate or a set of predicates would be precise enough for the verification of a given property. This oracle can be either how much the predicates 
contribute to the transition relation, i.e., how many transitions that are infeasible in the original system become feasible in the abstract, or actual verification being performed on a smaller instance of the same problem. In either case the analysis is property-directed. In the former case one identifies the relevant transitions for a given property and evaluates predicates wrt imprecision they introduce. }

\subsection{Incremental Abstraction}
\label{sec:threory}
Our goal is to assess imprecision of a set of predicates by building on the imprecision of the subsets, i.e., if at least one of the predicate sets $\varphi_1$ and $\varphi_2$ is not {\em precise}, we would like to be able to decide if combining the two sets, $\varphi_1 \cup \varphi_2$, will incur at least the same level of imprecision or not. So we present some definitions below that will be used for imprecision assessment. 

\begin{definition}[Orthogonal abstractions]
Two abstraction functions $\alpha_1$ and $\alpha_2$ are orthogonal if $\alpha_1 \circ \alpha_2 = \alpha_2 \circ \alpha_1$.
\end{definition}

\begin{definition}[Incremental abstraction]
An abstraction $\alpha$ is incremental if it can be built in terms of orthogonal abstractions $\alpha_1$ and $\alpha_2$, i.e., $\alpha=\alpha_1 \circ \alpha_2 = \alpha_2 \circ \alpha_1$, and satisfy the property that for every concrete state $s$. $\alpha_2(s) \sqsubseteq \gamma_1(\alpha(s))$ and $\alpha_1(s) \sqsubseteq \gamma_2(\alpha(s))$.
\end{definition}

\begin{definition}[Disjoint abstractions]
Two predicate abstraction functions $\alpha_1$ and $\alpha_2$ are called {\em disjoint} if they are defined over predicates whose scopes are disjoint, i.e., $V(\varphi_1) \cap V(\varphi_2) = \emptyset$, where $(\varphi_1,b_1)$ and $(\varphi_2,b_2)$ denote the predicate sets and the boolean variables that define $\alpha_1$ and $\alpha_2$, respectively. 
\end{definition}

Predicate abstraction on disjoint predicate sets yield orthogonal abstractions, which can be used to construct incremental abstractions:
\begin{lemma}
\label{lemma:incr}
Given disjoint predicate abstraction functions $\alpha_1$ and $\alpha_2$ defined over $(\varphi_1,b_1)$ and $(\varphi_2,b_2)$, respectively, abstraction function $\alpha_3$ defined over $(\varphi_1 \cup \varphi_2,b_1 \cup b_2)$ can be defined incrementally.
\begin{proof} . i.e., $\alpha_3 = \alpha_2 \circ \alpha_1=\alpha_1 \circ \alpha_2$ follows from the fact that $\alpha_1$ and $\alpha_2$ are disjoint:
\begin{equation*}
\begin{split}
 & = \alpha_2 \circ \alpha_1(s) \\
                   & = \exists V(\varphi_2). (\exists V(\varphi_1). s \wedge \bigwedge_{i=1}^{|\varphi_1|} \varphi_{1,}i \iff b_{1,i}) \wedge \bigwedge_{i=1}^{|\varphi_2|} \varphi_{2,i} \iff b_{2,i}\\
                    & = \exists V(\varphi_2). (\exists V(\varphi_1). s \wedge \bigwedge_{i=1}^{|\varphi_1|} \varphi_{1,}i \iff b_{1,i} \wedge \bigwedge_{i=1}^{|\varphi_2|} \varphi_{2,i} \iff b_{2,i})\\
                    & = \exists V(\varphi_2). V(\varphi_1). (s \wedge \bigwedge_{i=1}^{|\varphi_1|} \varphi_{1,}i \iff b_{1,i} \wedge \bigwedge_{i=1}^{|\varphi_2|} \varphi_{2,i} \iff b_{2,i})\\
                     & = \alpha_3(s)
\end{split}
\end{equation*}
Showing case $\alpha_3 = \alpha_1 \circ \alpha_2$ is similar. $\alpha_1(s) \sqsubseteq \gamma_2(\alpha_3(s))$ ($\alpha_2(s) \sqsubseteq \gamma_1(\alpha_3(s))$) also follows from the fact that $\alpha_1$ and $\alpha_2$ are disjoint and existential abstractions, and, hence, the new predicates on the new variables in $\alpha_2$ ($\alpha_1$) will introduce new behaviors as they were kept in their concrete domain when $\alpha_1$ ($\alpha_2$) was applied. 
\end{proof}
\end{lemma}

We would like to define a notion of imprecision that can be utilized in determining whether an abstraction may yield spurious behavior. The basic intuition in our formulation is that if an  
abstraction makes a transition to have a weaker precondition as to enable triggering more transitions  or the same transitions but in additional ways in backward-image computation compared to what was possible in the concrete case then we consider that abstraction as imprecise. The formal definition follows.

\begin{definition}[Transition-level Imprecision]
\label{def:trlimp}
Let  $T$ denote a transition system $T=(S,I,R)$ and  $\alpha$ denote an abstraction function.  
$\alpha$ is imprecise wrt transitions $r_1, r_2 \in R$ if 
$r_2 \wedge Range(Pre[r_1](true))  \sqsubset \gamma(\alpha(r_2) \wedge Range(Pre[\alpha(r_1)](true)))$ or 
$r_1 \wedge Range(Pre[r_2](true))  \sqsubset \gamma(\alpha(r_1) \wedge Range(Pre[\alpha(r_2)](true)))$, where $Range(f)$ rewrites the formula $f$ by renaming variables with their next state versions.
\end{definition}

\paragraph{Remark} It is important to note that Definition \ref{def:trlimp} points out to imprecision through the $\sqsubset$ operator, i.e., abstract version of transition $r_2$ is enabled from states that can be reached via executing the abstract version of $r_1$ in ways that were not possible in the concrete transition system $T$. An example is provided in Section \ref{sec:onsame}. 

The following lemma states that imprecise abstractions carry their imprecision in incremental abstractions. 

\begin{lemma}
\label{lemma:transImpr}
Let $T$ denote a transition system $T=(S,I,R)$ and $\alpha_1$ and $\alpha_2$ denote disjoint  abstraction functions. If $\alpha_1$ is imprecise wrt transitions $r_1, r_2 \in R$ then 
both $\alpha_1 \circ \alpha_2$ and $\alpha_2 \circ \alpha_1$ are imprecise wrt transitions $r_1, r_2$. 
\begin{proof}
Follows from 1) Lemma \ref{lemma:incr}, 2) predicate abstraction yielding an over-approximate pre-image operator, i.e., $Pre[\alpha_1(R)](true) \sqsubseteq \gamma(Pre[\alpha_2 \circ \alpha_1(R)](true)])$, 3) monotonic nature of the pre-image computation, and 4) $ a \sqsubset b$ and $b \sqsubseteq c$ implies $a \sqsubset c$.
\end{proof}
\end{lemma}

\begin{lemma}[Preservation for ECTL]
\label{lemma:ectl}
Given an existential abstraction function $\alpha$, an ECTL\footnote{ECTL is a fragment of CTL that involves temporal operators that are preceded by existential quantification, $E$, over states.} formula $\phi$, and transitions systems $T_1$ and $T_2$ such that $T_1 \sqsubseteq_{\alpha} T_2$, $T_1 \models \phi \implies T_2 \models \alpha(\phi)$. 
\begin{proof}
Follows from the fact that existential abstraction defines a simulation relation between $T_1$ and $T_2$ and preserves transitions, and, hence, paths in the abstracted system. 
\end{proof}
\end{lemma}

Lemma \ref{lemma:ectl} states that if a transition system satisfies an ECTL property then the existential abstraction of the abstracted transition system also satisfies this property.

\begin{lemma}
\label{lemma:incrAbs}
Let $T$ denote a transition system $T=(S,I,R)$ and $\alpha_1$ and $\alpha_2$ denote orthogonal abstraction functions. If $\alpha_1(T)$  does not satisfy an ACTL property $\phi$ then  neither $\alpha_1 \circ \alpha_2(T)$ nor $\alpha_2 \circ \alpha_1(T)$ satisfy $\phi$.
\begin{proof}
Follows from the fact that existential abstraction preserves transitions between mapped states and, hence, preserves paths. If an ACTL property is not satisfied by $\alpha_1(T)$ then 
it satisfies the negation, which is an ECTL property. From Lemma \ref{lemma:ectl} it follows that both $\alpha_1 \circ \alpha_2(T)$ and $\alpha_2 \circ \alpha_1(T)$ will 
contain the infeasible counter-example path.
\end{proof}
\end{lemma}

Lemma \ref{lemma:incrAbs} implies that if we are aware of an abstraction $\alpha$ that is imprecise for analyzing a transition system $T$ for a given ACTL property $\phi$, i.e., the property is not satisfied and the counter-example is infeasible,  incremental abstractions that involve $\alpha$ will also be imprecise for verifying $\phi$.

\subsection{Inferring Predicates from Transition-Level Imprecision}
\label{sec:onsame}

In this section, we present our first approach to choosing predicates, which is based on transition-level imprecision of predicates and builds upon Lemma \ref{lemma:transImpr}. We compute imprecision of predicates wrt transition pairs 
for each predicate as well as for each pairwise combination of the predicates using algorithm {\bf CompTransLevelImp} that is given in Figure \ref{fig:computetrl}. We quantify imprecision in terms of the number of additional predicate regions covered in enabling another transition for backward-image computation. When the imprecision is computed for an individual predicate (lines 8-17) the imprecision per transition pair can be 2 at maximum and when it is computed for a pair of predicates (lines 18-28) the same metric can be 4 at maximum.  
It should be noted that imprecision values for each pair of transition is summed up to compute the actual precision score for individuals (line 16) as well as for pairs of predicates (line 26).

\begin{figure}[th!]
\centering
\includegraphics[width=0.8cm]{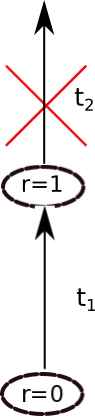}
\caption{Concrete transition $t_2$ cannot be executed from a state that is reached via $t_1$.}
\label{fig:concrete}
\end{figure}

\begin{figure}[th!]
\centering
\includegraphics[width=5cm]{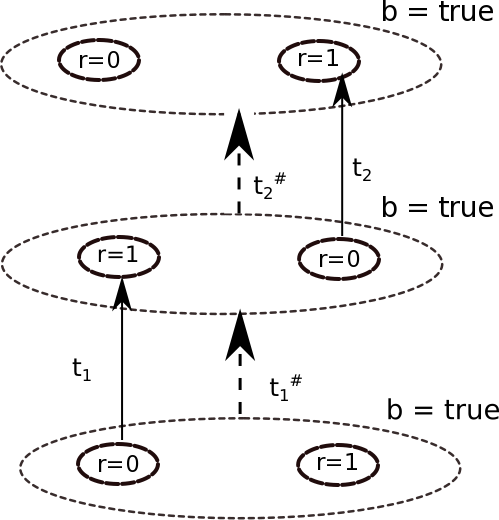}
\caption{Abstracting $t_1$ and $t_2$ with respect to predicate set $\{r\leq 1\}$.}
\label{fig:abs1}
\end{figure}

\begin{figure}[th!]
\centering
\includegraphics[width=3cm]{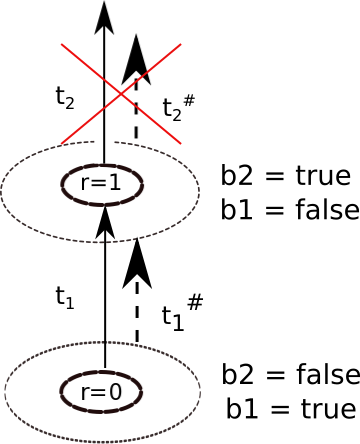}
\caption{Abstracting $t_1$ and $t_2$ with respect to predicate set $\{r=0, r=1\}$.}
\label{fig:abs2}
\end{figure}

As an example, consider the following transitions: $t_1 \equiv pc_1=a \wedge r=0 \wedge r'=r+1 \wedge pc_1'=c$ and  $t_2 \equiv pc_2=a \wedge r=0 \wedge r'=r+1 \wedge pc_2'=c$.  In the concrete semantics, $t_1 \wedge Range(pre[t_2](true))$ yields false as illustrated in Figure \ref{fig:concrete} as 
there is no state from which we can execute $t_1$ and then $t_2$. If we use predicate abstraction while using the predicate set $\{r\leq1\}$ and boolean variable $b$, we get the following abstract versions: $\alpha(t_1) \equiv pc_1=a \wedge b \wedge b'  \wedge pc_1'=c$ and  $\alpha(t_2) \equiv pc_2=a \wedge b \wedge b' \wedge pc_2'=c$. Now, if we compute 
$\alpha(t_1) \wedge Range((pre[\alpha(t_2)](true))$, the result will not be false as it evaluates to $pc_1=a \wedge b \wedge b' \wedge pc'_2=a \wedge b'$. As shown in Figure \ref{fig:abs1}, this means that abstract transition $t^{\sharp}_2$ is enabled from a state that is reached  executing abstract transition $t^{\sharp}_1$.
The way we quantify this extra behavior introduced by the abstraction is the number of extra predicate regions or cubes that enables the triggering of $t_1$ by $t_2$ when computing the image backward.  In this case the additional region covered when triggering $t_1$ by $t_2$ is when the predicate evaluates to true, e.g, $r\leq1$ or $b=true$. So the score for imprecision is 1 in this case. 
Figure \ref{fig:abs2} shows a case in which the predicate set $\{r=0,r=1\}$ avoids the imprecision that was possible with predicate set $\{r \leq 1\}$. This is because concrete state $r=0$ and concrete state $r=1$ are mapped to separate abstract states.

\begin{figure}[th!]
\begin{algorithmic}[1]
\State {\bf CompTransLevelImp}($T=(S,I,R)$: Transition System,$\varphi$: Set of predicates)
\State {\em Output:} $IS:  \varphi \to \mathcal{Z}$: Individual Imprecision Scores, $PwS: \varphi \times \varphi \to \mathcal{Z}$: Pairwise Imprecision Scores
\State $IS \gets \lambda x. 0$ 
\State $PwS \gets \lambda x.y. 0$
\For{each $r_j \in R$}
  \For{each $r_k \in R$}
    \For{each $\varphi_i \in \varphi$}
        \State Let $\alpha_i$ pred. abs. using $(\{\varphi_i,\},\{b_i\})$
        \State Let $Cube_{conc}$ denote the set $\{ \varphi_i , \neg \varphi_i\}$
         \State $conc \gets r_k \wedge Range((pre[r_j](true)))$ 
         \State $concCov \gets |\{ cube |  cube \in Cube_{conc}  \text{ and } conc \wedge cube \not = false \}|$
         \If{$concCov <2$}
            \State $abs \gets \alpha_i(r_k) \wedge Range((pre[\alpha_i(r_j)](true))$
            \State  Let $Cube_{abs}$ denote the set $\{ b_i , \neg b_i\}$
             \State $absCov \gets |\{ cube |  cube \in Cube_{abs} \text{ and } abs \wedge cube \not = false\}$
            \State $IS \gets IS[\varphi_i \mapsto IS[\varphi_i] + absCov - concCov]$              
         \EndIf
         \For{each $\varphi_m \in \varphi, \varphi_m \not = \varphi_i$}
                \State Let $\alpha_c$ pred. abs. using $(\{\varphi_i,\varphi_m\},\{b_i,b_m\})$
                \State Let $Cube_{conc}$ denote the set $\{ \varphi_i \wedge \varphi_m,  \varphi_i \wedge \neg \varphi_m, \neg \varphi_i \wedge \varphi_m, \neg \varphi_i \wedge \neg \varphi_m\}$
                \State $concCov \gets |\{ cube |  cube \in Cube_{conc}  \text{ and } conc \wedge cube \not = false \}|$
                \If{$concCov < 4$}           
                    \State  $abs \gets \alpha_c(r_k) \wedge Range(pre[\alpha_c(r_j)](true))$   
                    \State  Let $Cube_{abs}$ denote the set $\{ b_i \wedge b_m, b_i \wedge \neg b_m, \neg b_i \wedge b_m, \neg b_i \wedge \neg b_m\}$
                    \State $absCov \gets |\{ cube |  cube \in Cube_{abs} \text{ and } abs \wedge cube \not = false\}$
                    \State $PwS \gets PwS[(\varphi_i,\varphi_m) \mapsto PwS(\varphi_i,\varphi_m) + absCov - concCov]$
                \EndIf
         \EndFor
  \EndFor
 \EndFor  
\EndFor
\end{algorithmic}
\caption{An algorithm for computing transition level imprecision of predicates where  $IS$ stores the imprecision scores of individual predicates and $PwS$ stores the imprecision scores of pairwise combinations.}
\label{fig:computetrl}
\end{figure}
 
Once the imprecision scores are computed, the next step is to consider all feasible configurations by considering Lemma \ref{lemma:transImpr}, which states that when disjoint abstractions are combined the imprecision of the the individual abstractions will be carried to the new abstraction, which should be avoided. So algorithm {\bf ChoosePredsTransLevelImp}, given in Figure \ref{fig:choosetrl}, calls algorithm {\bf ExploreConfigTransLevelImp}, given in Figure \ref{fig:exploretrl}, to compute all feasible combinations of predicates up to a given bound and returns the configuration that yields the smallest imprecision score. It breaks any ties in choosing those with the maximum number of variables. Any ties at this level will be broken by choosing the one with the minimum number of predicates.

\begin{figure}[th!]
\begin{algorithmic}[1]
\State {\bf ChoosePredsTransLevelImp}($\varphi$: Set of Predicates, $IS:  \varphi \to \mathcal{Z}$: Individual Imprecision Scores, $PwS: \varphi \times \varphi \to \mathcal{Z}$: Pairwise Imprecision Scores, $k: \mathcal{N}$: Depth Bound)
\State // (predicate set, number of variables, imprecision score)
\State global $best: (\mathcal{N} \to \mathcal{P}(\varphi),\mathcal{N}, \mathcal{N})$, $best \gets (\lambda x. \emptyset,0,\infty)$
\State {\bf ExploreConfigTransLevelImp}($\emptyset$, 0, k)
\State {\bf return} $best[i]$, $ 1 \leq i \leq k$ with {\bf MIN} imprecision score, break ties with {\bf LARGEST} \#of variables, {\bf SMALLEST} \# of predicates
\end{algorithmic}
\caption{An algorithm for choosing a combination of predicates with the smallest imprecision score, maximum number of variables, and minimum number of predicates.}
\label{fig:choosetrl}
\end{figure}

Algorithm {\bf ExploreConfigTransLevelImp} excludes predicates that have non-zero imprecision scores at the individual level (line 5). It also excludes predicates that share variables with such predicates (lines 2 and 5).  However, pairwise imprecision scores for the predicates in a configuration is summed up to quantify the imprecision of that configuration. For each level in the configuration tree, whose height equals to the provided depth bound, 
the best solution encountered so far is recorded in a global triple $best$\footnote{We use the syntax $triple.two$ and $triple.three$ to access the second and the third items in the triple.} and updated whenever a better configuration is generated. It is important to note that ties are broken in the same order used in the global configuration: imprecision score followed by the number of variables. 

\begin{figure}[th!]
\begin{algorithmic}[1]
\State {\bf ExploreConfigTransLevelImp}($curSol: \mathcal{P}(\varphi)$, $level: \mathcal{N}$: Current Depth, $k: \mathcal{N}$: Depth Bound)
\State Let $excludeVars \gets \{ var \ | \ var \in Scope(\varphi_i) \wedge \varphi_i \in \varphi \wedge IS[\varphi_i] > 0\}$
\If{$level < k$}
  \For{each $\varphi_i \in \varphi$ s.t. $\varphi_i \not \in curSol$}
     \If{$curSol \cup \{\varphi_i\} \not \in visited$ and $IS[\varphi_i] = 0$ and $Scope(\varphi_i) \cap excludeVars = \emptyset$}
          \State $curSol \gets curSol \cup \{\varphi_i\}$
           \State $visited \gets visited \cup \{curSol\}$
          \State $impScore \gets$ $ \sum_{\varphi_j, \varphi_k \in curSol, \varphi_j \not = \varphi_k} PwS(\varphi_j, \varphi_k)$
          \State $numVars \gets |\bigcup_{\varphi_k \in curSol} Scope(\varphi_k)|$
          \If{($impScore < best[level+1].three$ ) or \State ($impScore = best[level+1].three$ and \State $numVars > best[level+1].two$}
              \State $best[level+1] \gets (curSol, numVars, impScore)$
          \EndIf
              \State {\bf ExploreConfigTransLevelImp}($curSol$, $level+1$, $k$)
              \State $curSol \gets curSol \setminus \{[\varphi_i\}$          
     \EndIf
  \EndFor
\EndIf
\end{algorithmic}
\caption{An algorithm for computing feasible combination of predicates by avoiding some possibilities of disjoint abstractions and computing the imprecision scores.}
\label{fig:exploretrl}
\end{figure}

\subsection{Inferring Predicates from Small Instance Verification}
\label{sec:small}

In this section, we present our second approach to choosing predicates, which is based on the actual verification results on a smaller instance of the transition system and builds on Lemma \ref{lemma:incrAbs}. 
When we say smaller instance we mean less number of concurrent components and assume that there exists a simulation relation between the smaller and the large instance, i.e., all the executions in the smaller instance are preserved in the large instance.
The  intuition behind this approach is that if an abstraction yields an infeasible counter-example path for the smaller instance then combining this abstraction with an orthogonal abstraction and using the incremental abstraction in the large instance will preserve the same infeasible counter-example path. So in the verification of the large instance, we should avoid configurations that are obtained using incremental abstractions where one of the abstractions is known to produce an inconclusive verification result in the smaller instance. 

\begin{figure}[th!]
\begin{algorithmic}[1]
\State {\bf ComputeCompatibility}($T=(S,I,R)$: Transition System,$\varphi$: Set of predicates, $\phi$: Correctness property in ACTL)
\State {\em Output} $compatibility: \varphi \times \varphi \rightarrow  \{0,1\}$: Pairwise Compatibility
\For{each $\varphi_i \in \varphi$}
   \For{each $\varphi_j \in \varphi$ s.t. $\varphi_i \not = \varphi_j$}
        \State $commonVars \gets Scope(\{\varphi_i,\varphi_j\}) \cap Scope(\phi)$
        \If{$commonVars \not = \emptyset$}
            \State $\varphi_{prop} \gets Predicates(\phi,commonVars)$
        \Else
            \State $\varphi_{prop} \gets \emptyset$
        \EndIf
        \State $\varphi_{try} \gets \{\varphi_i,\varphi_j\} \cup \varphi_{prop}$
        \State Let $\alpha_{try}$ defined using $(\varphi_{try},b_{try})$
        \State Let $T_1=(\alpha_{try}(S),\alpha_{try}(I),\alpha_{try}(R))$ 
        \If{$T_1 \models \llbracket \alpha_{try}(\phi) \rrbracket $}
            $compatibility \gets compatibility[(\varphi_i ,\varphi_j) \mapsto 1]$
        \Else
            $\ compatibility \gets compatibility[(\varphi_i ,\varphi_j) \mapsto 0]$   
        \EndIf
   \EndFor
\EndFor
\State {\bf return} $compatibility$
\end{algorithmic}
\caption{An algorithm for checking pairwise compatibility of predicates by applying the combined abstraction to a small instance, $T$, of a transition system.}
\label{fig:compat}
\end{figure}

Algorithm {\bf ComputeCompatibility} gets as input a small instance of the problem we would like to verify, the set of predicates, and an ACTL property representing the correctness property. It goes over every pair of predicate to generate a unique abstraction function (lines 5-12). As part of this process, it determines whether any extra predicate needs to be added by checking if any of the predicates involve variables that appear in the property. If so (line 6), it includes the predicates that appear in the property and  involve those variables (line 7) in the predicate set (line 11).

\begin{figure}[th!]
\begin{algorithmic}[1]
\State {\bf ChoosePredsCompatibility}($\varphi$: Set of Predicates, $Comp: \varphi \times \varphi \to \mathcal{Z}$: Pairwise Compatibility Scores, $k: \mathcal{N}$: Depth Bound)
\State // (predicate set, number of variables, compatibility score)
\State global $best: (\mathcal{N} \to \mathcal{P}(\varphi),\mathcal{N}, \mathcal{N})$, $best \gets (\lambda x. \emptyset,0,0)$
\State {\bf ExploreConfigCompatibility}($\emptyset$, 0, k)
\State {\bf return} $best[i]$, $ 1 \leq i \leq k$ with {\bf MAX} cohesion, break ties with {\bf LARGEST} \# of variables, {\bf SMALLEST} \# of predicates
\end{algorithmic}
\caption{Algorithm for choosing compatible predicates based on their compatibility in terms of producing a conclusive result in the small instance.}
\label{fig:choosecompat}
\end{figure}

Once we have the compatibility scores, we run algorithm {\bf ChoosePredsCompatibility} given in Figure \ref{fig:choosecompat} that calls algorithm {\bf ExploreConfigCompatibility} 
to enumerate all feasible configurations and compute their cohesion scores. By cohesion of a configuration we mean compatibility among members of the configuration. 
So the cohesion score of a configuration denotes  the number of compatible pairs. Among all feasible configurations algorithm {\bf ChoosePredsCompatibility} 
returns the one with maximum cohesion. If there is a tie, it is broken by choosing the one with the largest number of variables. For those with the same number of variables, the tie is broken by choosing the one with the smallest number of predicates.  

\begin{figure}[th!]
\begin{algorithmic}[1]
\State {\bf ExploreConfigCompatibility}($curSol: \mathcal{P}(\varphi)$, $level: \mathcal{N}$: Current Depth, $k: \mathcal{N}$: Depth Bound)
\If{$level < k$}
  \For{each $\varphi_i \in \varphi$ s.t. $\varphi_i \not \in curSol$}
     \If{$curSol \cup \{\varphi_i\} \not \in visited$}
        \State $visited \gets visited \cup \{curSol\}$
        \State // Check for the possibility of a disjoint abstraction when there is incompatibility
        \State $disjointAbs \gets false$
        \For{each $\varphi_j$ s.t. $compat(\varphi_i,\varphi_j) = 0$}
            \State $\varphi^1 \gets curSol \setminus \{\varphi_j\}$     
            \State $ \varphi^2 \gets \{\varphi_j,\varphi_j\}$         
            \If{$Scope(\varphi^1) \cap Scope(\varphi^2) = \emptyset$}
                \State $disjointAbs \gets true$
                \State break
            \EndIf
        \EndFor        
        \If{$disjointAbs=false$}
          \State $curSol \gets curSol \cup \{\varphi_i\}$
           \State $cohesion \gets \sum_{\varphi_j, \varphi_k \in curSol, \varphi_j \not = \varphi_k} compat(\varphi_j, \varphi_k)$  
            \State $numVars \gets |\bigcup_{\varphi_k \in curSol} Scope(\varphi_k)|$
          \If{($cohesion > best[level+1].three$) or \State ($cohesion = best[level+1].three$ and $numVars>best[level+1].two)$}
              \State $best[level+1] \gets (curSol, numVars, cohesion)$
          \EndIf
              \State {\bf ExploreConfigTransLevelImp}($curSol$, $level+1$, $k$)
              \State $curSol \gets curSol \setminus \{[\varphi_i\}$                              
        \EndIf
     \EndIf
  \EndFor
\EndIf     
\end{algorithmic}
\caption{An algorithm for enumerating all feasible configurations and computing their cohesion scores.}
\label{fig:explorecompat}
\end{figure}

Algorithm \ref{fig:explorecompat} uses Lemma \ref{lemma:incrAbs} to avoid adding a predicate $\varphi_i$ to the current solution set $curSol$ if this predicate is incompatible with one of the predicates $\varphi_j$ and the abstraction defined on $\{\varphi_i,\varphi_j\}$ and $curSol \setminus \{\varphi_j\}$ are disjoint abstractions (lines 7-16). It keeps a record of the best solution for each level in the exploration tree and 
updates the best solution in the global triple $best$ for the current level when a configuration has a higher cohesion. Ties are broken using the same scheme as in the global decision made in algorithm {\bf ChoosePredsCompatibility}: using the number of variables followed by the number of predicates.

%% file: experiments.tex
We have conducted experiments to evaluate the effectiveness of the proposed heuristics. 
The experiments have been executed on a 64-bit Intel Xeon(R) CPU with 8 GB RAM running Ubuntu 14.04 LTS. 
We used a model of Airport Ground Network Controller (AGNC) \cite{YB09} and a model of a character-special device driver. AGNTC is a resource sharing model for multiple processes, where the resources are taxiways and runways of an airport ground network and the processes are the arriving and departing airplanes. We changed the AGNTC model given in \cite{YB09} to obtain two variants  by 1) using one of the two mutual exclusion algorithms, the ticket algorithm \cite{And91} ({\tt airport4T}) and Lamport's Bakery algorithm ({\tt airport4B}), for synchronization on one of the taxiways, 2) making parked arriving airplanes fly and come back to faithfully include the mutual exclusion model, i.e., processes go back to $think$ state after they are done with the critical section in order to attempt to enter the critical section again, and 3) removed the departing airplanes. We verified three safety properties, {\tt p1-p3}, and one liveness property, {\tt p4}, for each of the variants. The character-special device driver is a pedagogical artifact from a graduate level course. It models two modes, where one of the modes allows an arbitrary number of processes to perform file operations concurrently whereas the other mode allows only one 
process at a time. The synchronization is performed using semaphores modeled with integer variables. The {\tt ioctl} function is used to change from one mode to another when there are no other processes working in the current mode. The total number of processes in each mode is kept track of to transition from one mode to another in a safe way. The update of the process counters are also achieved using semaphores modeled with integer variables. 

\begin{sidewaystable}
\centering
\begin{footnotesize}
\begin{tabular}{|r|r|r|r|r|r|r|r|r|r|r|r|r|r|}  \hline
{\bf Problem} & {\bf $|V|$} & \multicolumn{4}{c|}{\bf State Space} & \multicolumn{4}{c|}{\bf Initial States} & \multicolumn{4}{c|}{\bf Trans. Rel.} \\ 
{\bf Instance} &  & {\bf BDD} & {\bf Poly} & {\bf (G)EQ} & {\bf \#Dis} & {\bf BDD} & {\bf Poly} & {\bf (G)EQ} & {\bf \#Dis} & {\bf BDD} & {\bf Poly} & {\bf (G)EQ} & {\bf \#Dis} \\ \hline
airport2T & 9I, 8B& 11 & 1 & 3 & 1 & 12 & 1 & 4 & 1 & 527 & 10 & 75 & 10 \\ \hline
airport3T & 10I, 12B& 13 & 1 & 5 & 1 & 14 &1  &5  &1  &852  &8  & 44 & 8 \\ \hline
airport4T & 11I, 16B & 19 & 1 & 5 & 1 & 19 & 1& 5 & 1 & 1236 & 9 & 57 & 9 \\ \hline
airport2B & 7I , 8B & 11 & 1 & 5  & 1 & 12 & 1 & 5 & 1 & 542 & 13 & 73 & 11 \\ \hline
airport3B & 8I, 12B & 19 & 1 & 5& 1 & 20 & 1 & 5 & 1 &822  & 13 & 72 &11  \\ \hline
airport4B & 9I, 16B& 19 & 1 & 6 & 1 & 20 & 1 & 6 & 1 & 1298 & 23 & 163 & 15 \\ \hline
charDriver2 & 4I, 11B& 11 & 1 & 0 & 1 & 14 & 1 & 2 & 1 & 1097 & 13 & 32 & 13 \\ \hline
charDriver3 & 4I, 16B  & 22 & 1 & 0 & 1 & 21 &1  &2  &1  & 2267 & 13 & 32 & 13 \\ \hline
charDriver4 & 4I, 21B & 17 & 1& 0 & 1& 16 & 1 & 2 & 1 & 1476 & 13 & 32 & 13 \\ \hline
\end{tabular}
\end{footnotesize}
\caption{Sizes of the problem instances in terms of sizes of the state space restriction constraints, the initial states and the transition relation. {\bf BDD}, {\bf Poly}, {\bf (G)EQ}, and {\bf \#Dis} represent the size of the BDD, number of polyhedra, total number of equality and inequality constraints, and the number of disjuncts in the respective constraint, respectively.}
\label{table:size}
\end{sidewaystable}

Table \ref{table:size} shows sizes of the problem instances in terms of integer $I$ and boolean $B$ variables and sizes of the state space, the initial state, and the transition relation, which were demonstrated in terms of the Binary Decision Diagram (BDD) size for the boolean domain, number of polyhedra and number of integer constraints for the integer domain, and number of composite\footnote{A composite formula consists of conjunction of a boolean and integer formula.} disjuncts. 
ALV applies a simplification heuristic \cite{YB02c} to reduce size of the constraints and the data in Table  \ref{table:size} represents the values after simplification. Although {\tt airport4T} looks like to have a smaller integer constraint size than {\tt airport2T}, this is due to simplification. The number of equality constraints in the transition relation before simplification is is 1070 for {\tt airport2T}  and 2416 for {\tt airport4T}.

\begin{table*}[th!]
\centering
\begin{footnotesize}
\begin{tabular}{|r|r|r|r|r|r|r|r|r|r|r|r|} \hline
 & \multicolumn{2}{c|}{} & \multicolumn{9}{c|}{\bf PARTIAL PREDICATE ABSTRACTION} \\ \cline{4-12} 
{\bf Problem} &  \multicolumn{2}{c|}{\bf  POLY}  & \multicolumn{2}{c|}{\bf CEGAAR} &  & \multicolumn{3}{c|}{\bf TRLIMP} &  \multicolumn{3}{c|}{\bf  COMPAT} \\ \cline{2-12}
&Time & Mem  & Time & Mem & {\bf \#P} & In/Pw & Time & Mem & C/I & Time & Mem \\ \hline
air2T-P1 & 0.01 & 3.44& 0.31 & 6.42 & 15 & 4/46 & 0.01 & 8.70 & \multicolumn{3}{c|}{$NA$}  \\ \cline{1-5} \cline{8-12}
air3T-P1 & 0.02 & 5.96 & 0.63 & 9.13 & & & 2.14 & 21.23 & 105/0 & 2.14 & 21.23 \\ \cline{1-5} \cline{8-9} \cline{11-12}
air4T-P1 & 0.04 & 8.81   & 1.11 & 12.50 &  &  & 5.87 & 27.90 &  & 5.87 & 27.90 \\ \hline
air2T-P2 & 0.28 & 5.41  & 7.01 & 55.75 & 14  & 4/42 &  \multicolumn{2}{c|}{$UV$} & \multicolumn{3}{c|}{$NA$}\\ \cline{1-5} \cline{8-12}
air3T-P2 & 12.95 & 14.42 & 264.01 & 165.11 & & & \multicolumn{2}{c|}{$UV$} & 8/83 & 84.89 & 28.17 \\ \cline{1-5} \cline{8-9} \cline{11-12}
air4T-P2 & \multicolumn{2}{c|}{$TO$} & \multicolumn{2}{c|}{$TO$} &  & & \multicolumn{2}{c|}{$UV$} &  & 172.92 & 36.00 \\ \hline
air2T-P3 & 0.29 & 5.43 & 55.94 & 1963.37  & 15  & 4/46 & 0.06 & 8.70 & \multicolumn{3}{c|}{$NA$} \\ \cline{1-5} \cline{8-12}
air3T-P3 & 12.90 & 14.43 & 26.23 & 593.84 & & & 1.71 & 13.32 & 100/5 & 1.71 & 13.31 \\ \cline{1-5} \cline{8-9} \cline{11-12}
air4T-P3 & \multicolumn{2}{c|}{$TO$} & 50.44 & 974.97  &  & & 6.73 & 23.00 &  & 6.73 & 23.00 \\ \hline
air2T-P4 & 0.88 & 9.79   & 8.75 & 94.84 & 15 & 4/46 & 0.33 & 9.37 & \multicolumn{3}{c|}{$NA$}\\ \cline{1-5} \cline{8-12}
air3T-P4 & 13.12 & 63.21 & 59.04 & 33.30 & & & 2.65 & 26.96 &  61/43 & 2.65 & 26.96 \\ \cline{1-5} \cline{8-9} \cline{11-12}
air4T-P4 & \multicolumn{2}{c|}{$TO$}   & \multicolumn{2}{c|}{$TO$} &  &  & 17.23 & 39.32 & & 17.23 & 39.32 \\ \hline
air2B-P1 & 0.01 & 3.95  & 0.30 & 6.67 & 16 & 4/50 & 0.01 & 11.22 & \multicolumn{3}{c|}{$NA$}  \\ \cline{1-5} \cline{8-12}
air3B-P1 & 0.02 & 7.97 & 1.36 & 18.15  & & & 2.94 & 22.12 & 120/0 & 2.94 & 22.12 \\ \cline{1-5} \cline{8-9} \cline{11-12}
air4B-P1 & 0.06 &14.75   & 8.49 & 40.03 &  &  &16.80  & 43.80 &  & 16.80 & 43.80  \\ \hline
air2B-P2 & 1.44 & 6.47  & 11.95 & 110.41 & 15 & 4/46&  \multicolumn{2}{c|}{$UV$} & \multicolumn{3}{c|}{$NA$}\\ \cline{1-5} \cline{8-12}
air3B-P2 & 402.64 & 69.99 &\multicolumn{2}{c|}{$TO$}  & & & \multicolumn{2}{c|}{$UV$} & 8/97 & 84.66 & 43.78 \\ \cline{1-5} \cline{8-9} \cline{11-12}
air4B-P2 & \multicolumn{2}{c|}{$TO$}  & \multicolumn{2}{c|}{$TO$} &  &  & \multicolumn{2}{c|}{$UV$} &  & \multicolumn{2}{c|}{$TO$} \\ \hline
air2B-P3 & 0.12 & 4.66  & 1.31 & 19.83 & 16 & 4/50 & 0.01 & 10.20 & \multicolumn{3}{c|}{$NA$}  \\ \cline{1-5} \cline{8-12}
air3B-P3 & 64.01 & 29.11 & 4.17 & 20.03 & & & 9.82 & 24.85 & 114/6 & 9.82 & 24.85 \\ \cline{1-5} \cline{8-9} \cline{11-12}
air4B-P3 & \multicolumn{2}{c|}{$TO$}  & \multicolumn{2}{c|}{$TO$} &  &  & 94.29  & 39.42  &  & 94.29 & 39.42  \\ \hline
air2B-P4 & 1.07 & 10.23  & 16.59 & 569.86  & 16  & 4/50 & 0.16 & 10.25 & \multicolumn{3}{c|}{$NA$} \\ \cline{1-5} \cline{8-12}
air3B-P4 & 64.01 & 29.10 & 44.40 & 38.62 & & & 6.62 & 23.91 & 98/22 & 6.62 & 23.91 \\ \cline{1-5} \cline{8-9} \cline{11-12}
air4B-P4 & \multicolumn{2}{c|}{$TO$}  & \multicolumn{2}{c|}{$TO$} &  & &  113.43 & 107.88 &  & 113.43 & 107.88 \\ \hline
cdr2 & 1.49 & 14.46   & \multicolumn{2}{c|}{$TO$} & 10 & 2/13 &  \multicolumn{2}{c|}{$UV$}  & \multicolumn{3}{c|}{$NA$} \\ \cline{1-5} \cline{8-12}
cdr3 & 5.38 & 25.53 & \multicolumn{2}{c|}{$TO$}  & &  & \multicolumn{2}{c|}{$UV$}   & 3/42 & 7.08 & 53.43 \\ \cline{1-5} \cline{8-9} \cline{11-12}
cdr4 & \multicolumn{2}{c|}{$TO$}    & \multicolumn{2}{c|}{$TO$} &   &  &  \multicolumn{2}{c|}{$UV$}  &  & 13.01 & 71.85 \\ \hline
\end{tabular}
\end{footnotesize}
\caption{Comparison of verification results using polyhedra-based representation ({\bf POLY}), partial predicate abstraction with CEGAR ({\bf PCEGAR}), partial predicate abstraction with selection heuristic that uses transition-level imprecision ($TRLIMP$), and predicate abstraction  with selection heuristic that uses small instance results ({\bf  COMPAT}). $TO$ denotes a timeout of 20 minutes or more. $UV$ means unable to verify. $NA$ means not applicable. Time is given in seconds and memory is in MBs.}
\label{table:res}
\end{table*}

\begin{table*}
\centering
\begin{footnotesize}
\begin{tabular}{|c|c|c|c|c|c|c|c|c|c|c|c|c|c|c|} \hline& $p_{0}$ & $p_{1}$ & $p_{2}$ & $p_{3}$ & $p_{4}$ & $p_{5}$ & $p_{6}$ & $p_{7}$ & $p_{8}$ & $p_{9}$ & $p_{10}$ & $p_{11}$ & $p_{12}$ & $p_{13}$\\
$p_{0}$ & $\tick$ & $\tick$ &  &  &  &  & 2 & 2 & 2 &  &  &  &  & \\
$p_{1}$ & $\tick$ & $\tick$ & $\tick$ & $\tick$ & $\tick$ & $\tick$,2 & $\tick$,2 & $\tick$,2 & $\tick$,2 &  &  &  &  & \\
$p_{2}$ &  & $\tick$ & $\tick$ &  &  & 2 &  & 2 & 2 &  &  &  &  & \\
$p_{3}$ &  & $\tick$ &  & $\tick$ &  & 2 & 2 &  & 2 &  &  &  &  & \\
$p_{4}$ &  & $\tick$ &  &  & $\tick$ & 2 & 2 & 2 &  &  &  &  &  & \\
$p_{5}$ &  & $\tick$,2 & 2 & 2 & 2 & $\tick$ & 4 & 4 & 4 & 2 & 2 & 2 & 2 & 2\\
$p_{6}$ & 2 & $\tick$,2 &  & 2 & 2 & 4 & $\tick$ & 4 & 4 & 2 & 2 & 2 & 2 & 2\\
$p_{7}$ & 2 & $\tick$,2 & 2 &  & 2 & 4 & 4 & $\tick$ & 4 & 2 & 2 & 2 & 2 & 2\\
$p_{8}$ & 2 & $\tick$,2 & 2 & 2 &  & 4 & 4 & 4 & $\tick$ & 2 & 2 & 2 & 2 & 2\\
$p_{9}$ &  &  &  &  &  & 2 & 2 & 2 & 2 & $\tick$ &  &  &  & \\
$p_{10}$ &  &  &  &  &  & 2 & 2 & 2 & 2 &  & $\tick$ &  &  & \\
$p_{11}$ &  &  &  &  &  & 2 & 2 & 2 & 2 &  &  & $\tick$ &  & \\
$p_{12}$ &  &  &  &  &  & 2 & 2 & 2 & 2 &  &  &  & $\tick$ & \\
$p_{13}$ &  &  &  &  &  & 2 & 2 & 2 & 2 &  &  &  &  & $\tick$\\ \hline 
\end{tabular}
\end{footnotesize}
\caption{Merging results of verification in the small compatibility and transition-level imprecision results for {\tt airport2T-p2}. Existence and absence of a $\tick$ means conclusive and inconclusive verification in the small, respectively. Existence and absence of a score  means no imprecision and the level of imprecision based on transition level compatibility, respectively.}
\label{table:matrix}
\end{table*}

We have used ALV\footnote{The tool can be downloaded from \url{http://www.tuba.ece.ufl.edu/spin17.zip}}, which has been extended to implement partial predicate abstraction with and without CEGAAR, to run the experiments. We extracted the predicates from the initial state and state space restrictions, the transition guards, and the correctness property. 
We have used 2 process versions (2 airplanes for AGNC and 2 user processes for charDrv) as the small instances and 3 and 4 process versions as the large instances. 
Table \ref{table:res} shows the experimental results for verifying problem instances for 2, 3, and 4 concurrent processes using four approaches: 1) polyhedra-based representation, {\bf POLY}  (without predicate abstraction) and partial predicate abstraction with 2) CEGAAR, 3) predicate selection heuristic, {\bf TRLIMP}, that uses transition-level imprecision, and 4) predicate abstraction with predicate selection heuristic, {\bf COMPAT}, that uses verification results based on the smaller instance. For {\bf TRLIMP}, $In$ and $Pw$ denote the number of predicates  and the number of predicate pairs with non-zero individual imprecision scores, respectively. For {\bf COMPAT}, $C$ and $I$ denote the number of compatible and incompatible predicate pairs, respectively. Time is measured in seconds and includes the construction time, which includes computing the initial abstraction for partial predicate abstraction, the verification time, and refinement time for CEGAAR. 
We used a time limit of 40 minutes running the instances. So those that did not finish by that limit are represented with $TO$. The instances that could not provide a conclusive result are denoted with $UV$. Memory represents the total memory used and expressed in MB.

 As the results show {\bf POLY} and {\bf CEGAAR} are generally effective in small size problems. For the {\tt charDriver} instances, {\bf CEGAAR} was not effective even for the smallest size. 
 Considering the largest instances (concurrent component of size 4), {\bf COMPAT} demonstrated the best performance. 
 {\bf COMPAT} was able to find precise combinations for all cases whereas {\bf TRLIMP} missed a property for each benchmark. 
 For the cases that {\bf TRLIMP} found a precise combination of predicates, it found the exact set as found by  {\bf COMPAT}.
 All instances except {\tt charDriver} instances could be verified by selecting 2 predicates, the depth bound in Algorithms \ref{fig:choosetrl} and \ref{fig:choosecompat}. For {\tt charDriver}, a predicate set of size 2 did time out. Using a predicate set of size 4 provided the results reported in the table. 


For computing compatibility,  verifying the small instance on a pair of predicates on average took under 2 secs for the {\tt airport} problems and 86.25 secs for the {\tt charDriver} problem, respectively. 
However, since computing transition-level imprecision scores on 4 processes did not scale, we used the 2 process versions, which on average took 62.75 secs for {\tt airportTicket}, and 25.85 secs for {\tt airportBakery} and 259.54 secs for {\tt charDrv} instances and used these to generate the predicate solutions that were used in column {\bf TRLIMP} in Table \ref{table:res}.

We investigated the root cause of {\bf TRLIMP} not being able to find a precise combination for the three problem instances. Table \ref{table:matrix} shows a combination of 
inferences made for pairs of predicates by {\bf TRLIMP} and {\bf COMPAT} for {\tt airport2T-p2}. We represented pairwise compatibility (incompatibility) based on small instance verification results with the existence (absence) of a $\tick$ symbol and pairwise imprecision (precision) score with existence (absence) of a positive score. In this case, the set of predicates suggested by {\bf TRLIMP} is $\{AirplaneA.0.a = main.0.t, main.0.s \geq AirplaneA.1.a\}$ ($\{p_{10},p_{13}\}$) and that of {\bf COMPAT} is $\{main.0.numRW16R = 0,main.0.numC3 = 0\}$ ($\{p_1,p_1\}$). As can be confirmed by Table \ref{table:matrix}, {\bf TRLIMP} infers that the predicates in {\bf COMPAT}'s solution set, $p_0$ and $p_1$, are compatible. So in its exploration, it considers these predicates as a candidate solution. However, it also decides that there is no imprecision due to the pair $(p_{10},p_{13})$, which in reality, i.e., small model based results, are not compatible. The reason  {\bf TRLIMP} prefers  $\{p_{10},p_{13}\}$ over $\{p_1,p_1\}$ is the number of variables abstracted (4)  is higher in the former than that (2) in the latter. Although  {\bf TRLIMP} can be configured on whether to consider the number of variables abstracted, we believe that often the number of variables can be critical as demonstrated in the {\tt charDriver} benchmark, i.e., abstracting a small number of variables did not provide the necessary state space reduction. 

\begin{table}[th!]
\centering
\begin{footnotesize}
\begin{tabular}{|r|r|r|} \hline
{\bf Problem} & {\bf Precision} & {\bf Recall} \\ \hline
airport2T-P1 & 50.00\% & 42.99\%\\
airport2T-P2 & 54.54\% & 43.63\%\\
airport2T-P3 & 50.00\% & 42.99\% \\
airport2T-P4 & 47.83\% & 50.00\%\\
airport2B-P1 & 52.00\%& 41.60\% \\
airport2B-P2 & 59.38\% & 42.86\%\\
airport2B-P3 & 52\% & 41.60\% \\
airport2B-P4 & 32\% &72.72\% \\
charDriver2 & 58.33\% & 23.73\\ \hline 
\end{tabular}
\end{footnotesize}
\caption{Precision and Recall values for predicting transition level incompatibility of predicates.}
\label{table:precRecall}
\end{table}

Table \ref{table:precRecall} presents precision and recall values for {\bf TRLIMP}. Precision has been computed as $\frac{|correctImprecisePairs|}{|imprecisePairs|}$, where \\ $|imprecisePairs|$ denotes the number of  pairs that have positive imprecision scores and $|correctImprecisePairs|$ denotes the number of pairs that have both positive imprecision scores and produce inconclusive verification results. Recall has been computed as $\frac{|correctImprecisePairs|}{|allImprecise|}$, where $|allImprecise|$ denotes the number of pairs that produce inconclusive verification results. As the numbers suggest, {\bf TRLIMP}, as a technique oblivious to the verified property, is on average 50\%  accurate in its identification of imprecise pairs, which could still produce conclusive verification results for most of the problems we used in our experiments. 
 
 As the experimental results suggest both  {\bf TRLIMP} and {\bf COMPAT} have potential in automated abstraction generation for the partial predicate abstraction technique. 
 The advantage of {\bf TRLIMP} is that, in principle, it does not require the problem instance to have concurrent components. However, as the problem size grows computing pairwise imprecision scores does not scale and individual imprecision scores may be the only data available assuming there is no smaller version.
 Its disadvantage, however, is not involving the correctness property in the computation of imprecision scores. 
 The advantage of {\bf COMPAT} is being truly property-directed and its main restriction is requiring the problem to have concurrent components that may be instantiated a number of times. 
 However, {\bf COMPAT} can be used in an incremental model development process as long as each new version adds new behaviors without removing any behaviors that existed in the old version.

%% file: relWork.tex
Variable dependency graphs in \cite{BS93,LA99,Kur02,CGJ03}  are used to iteratively infer variables needed to refine the abstraction. 
 In our approach we measure variable interactions in terms of the imprecision induced by individual as well as pairs of predicates and in terms of the compatibility of the predicates they appear in  the verification of small instances of a problem.
 Eliminating  predicates due to their redundancy are studied in \cite{CGT03} to improve efficiency of predicate abstraction. Our predicate elimination is also concerned with efficiency but we also consider precision in our decision.
\cite{JIG06} reports the improved precision of predicate abstraction on the strengthened  transition relation using automatically generated invariants. Our approach achieves a precise set of predicates by considering the predicates that already exist in the model. Using the predicates that appear in the guards of the transitions \cite{GS97} is the most basic approach to forming candidate predicate sets. We also use the predicates that appear in the correctness property and in the initial state and state space restriction specifications. 

Selecting predicates in the context of CEGAR has been studied in \cite{BLW15}. \cite{BLW15}  chooses among possible refinement schemes using a number of heuristics such as size of the variables' domain, the deepness of the pivot location, the length of the sliced infeasible prefixes, and how long the analysis needs to track additional information. Our selection heuristics are biased towards precision rather than efficiency although we try to maximize the number of abstracted variables and minimize the number of predicates. 
\cite{MS14} synthesizes predicates to improve precision of the numerical abstract domain to recover from imprecisions incurred by convex approximations. 
\cite{BLN13} reuses abstraction precision, e.g., the set of predicates used, in the model checking of new versions of the software. 

The small model theory we use in this paper differs from those used for verification of parameterized systems \cite{AHH13}. In the context of parameterized systems,  a small configuration of the system is used to come up with an abstract model. When such an abstraction cannot be shown to satisfy the property, the model is refined by considering a larger configuration. So verification in the small model is generalized to verification in the larger. In our setting, a larger model acts as an abstraction of the small model as all the behaviors in the small model are preserved while adding new behaviors. So falsification in the small is generalized to falsification in the larger leading to elimination of certain predicate combinations in the abstraction of the larger model.

%% file: conclusions.tex
\ignoreme{
State-explosion is an inherent problem in model checking. The abstract interpretation framework provides a systematic way of combining abstractions and approximations to push the limits of automated verification, specifically model checking. In this paper we have combined two  techniques, predicate abstraction and approximate fixpoint computations, that have been effective in the verification of infinite-state systems. This hybrid approach proves to be advantageous when the fixpoint computations do not converge for a given correctness property. In such cases letting part of the state space that causes non-convergence to be handled by fixpoint approximations and the rest of the state space abstracted using predicate abstraction can improve the verification time and in some cases the memory requirements. 
}

We have proposed two heuristics to choose predicates for partial predicate abstraction so  that the achieved state-space reduction through partial abstraction does not yield inconclusive 
verification results. We have formulated a notion of imprecision at the transition-level and a notion of compatibility among predicates based on small instance verification. 
Our heuristics are based on the aspect of incremental abstraction inheriting imprecision from its component abstractions. 
We leverage this theoretical result to soundly eliminate predicate combinations in our quest for a precise abstraction. 
Experimental results show that both heuristics have potential in automated abstraction. The main trade-off between the two heuristics relates to being property directed versus requiring to have a model structure that can be expanded with more functionality without losing any of  the existing ones.

For future work, we would like to investigate how to incorporate the correctness property into the transition-level imprecision computation and improve efficiency of computing the imprecision scores as well as compatibility measurement by reusing computation across checking different properties. Another direction we are interested in exploring is  incorporating imprecision inferring heuristics to the counter-example guided abstraction refinement process to eliminate candidate predicates that may potentially introduce imprecision. We would also like to apply these heuristics in the context of software model checking, which will provide better access to a large set of benchmarks. 